\documentclass[
reprint,amsmath,amssymb,aps,pra,floatfix,longbibliography]{revtex4-1}
\usepackage{graphicx}	
\usepackage{dcolumn}	
\usepackage{bm}            
\usepackage{color}

\begin{document}
\title{
A 
large
nonmagnetic
aluminum-and-Viton
vacuum chamber 
suitable for vacuum 
pressures down into the
extremely-high-vacuum 
(XHV)
regime
}

\author{N.T. McCall}
\author{H.-M. Yau}
\author{Z. Corriveau}
\author{J. Perez Garcia}
\author{T. Chauhan}
\author{G.K. Koyanagi}
\author{R.L. Lambo}
\author{M.C. George}
\author{C.H. Storry}
\author{E.A. Hessels} 
\email{hessels@yorku.ca}
\affiliation{Department of Physics and Astronomy, York University, Toronto,
Ontario M3J 1P3, Canada}

\date{\today} 

\begin{abstract}

A recently constructed large
nonmagnetic
vacuum chamber is described
and characterized.
The vacuum chamber was 
machined from 
19-inch 
cubes of 
aluminum and are  
sealed using 
dual 
Viton
O-rings
(with the 
interspace between 
the 
O-rings
evacuated by a 
separate vacuum 
pump).
17-inch-diameter
openings on all 
six sides 
allow for 
easy 
access to the 
experimental 
apparatus 
and 
for vacuum 
pumps with very 
large pumping 
speeds.
These 
17-inch 
holes were bored 
straight through 
the aluminum cube,
resulting in an 
interior volume 
of 
100~L
(almost 
90\% 
of the total volume
of the 
19-inch 
cube).
The 
17-inch
openings are sealed with 
19-inch-diameter
aluminum flanges.
The system is 
modular in that 
the cubes can be 
connected to one another
to form larger 
vacuum systems.
The chamber is designed
to allow for vacuum 
pressures 
of
$10^{-12}$~torr.
In many respects
(including nonmagnetic
properties, 
outgassing 
rate
after 
low-temperature 
baking,
weight, 
cost,
ease of machining,
and 
ease of baking due to 
the chamber's 
high thermal conductivity)
this system is 
superior to standard 
stainless-steel 
metal-sealed
CF
vacuum systems
and is highly recommended
for use by others.


\end{abstract}

\pacs{Valid PACS appear here}
\maketitle

\section{introduction}

\begin{figure}[b!]
\includegraphics[width=0.8\linewidth]
{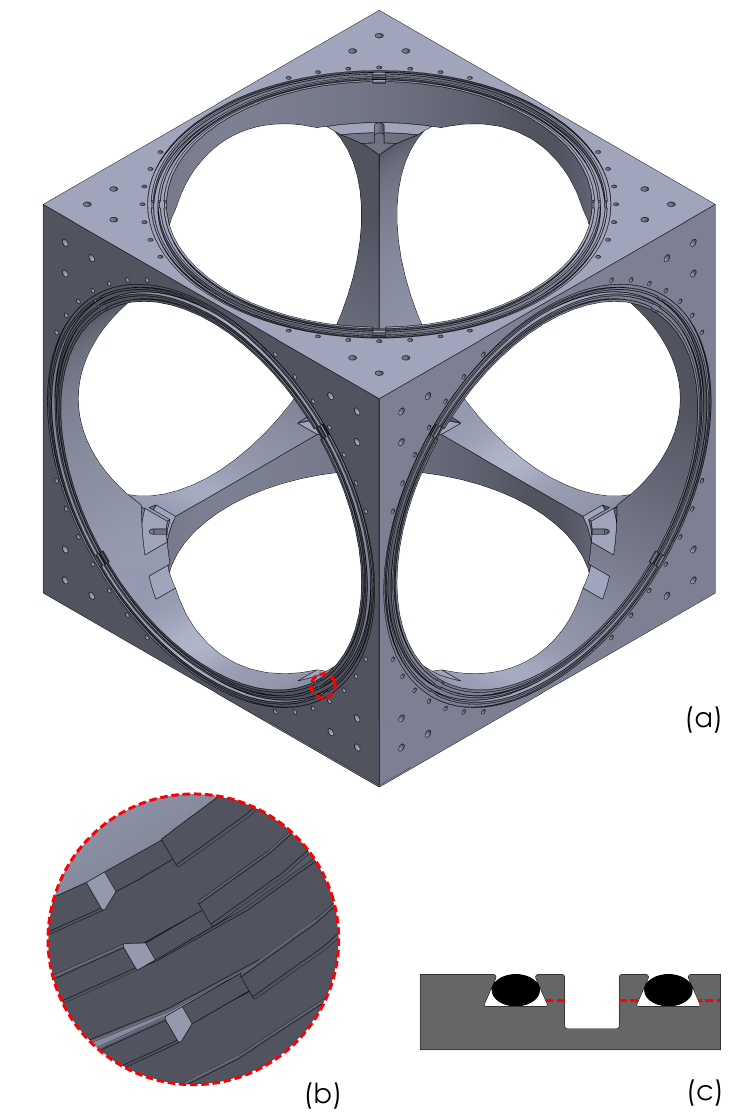}
\caption{
(color online)
One of the cubes that
makes up our vacuum system
is shown in panel 
(a).
The 
dual-O-ring
seal is expanded
in 
panels~(b)
and 
(c).
These
expanded views 
show the
half-inch-wide
cross-over 
vents
(red dashed lines in (c))
that 
pump out the
blind areas below the 
O-rings
to the intermediate vacuum 
(center channel in (c)),
or to the main vacuum
(to the right in (c)).
These vents are located 
at four locations around 
each opening,
one of which is
highlighted 
by the dashed red
circle in 
(a).
}
\label{fig:cubeDrawing}
\end{figure}

In this paper we describe
an 
aluminum-and-Viton
vacuum system
that is suitable for
the
extremely-high-vacuum 
(XHV)
regime.
The vacuum chamber
is machined
(by
Attica
Manufacturing)
from 
19-inch
aluminum cubes
(see 
Fig.~\ref{fig:cubeDrawing}(a)).
A 
dual-O-ring
design 
overcomes concerns about
scratches on the aluminum
sealing surfaces 
and 
permeation of gasses 
through the 
Viton
O-rings.
The cubes 
have
17-inch-diameter
openings on all 
six sides to allow 
for maximum possible 
access to the interior
volume
(of 
100~L)
and
for maximal possible
pumping speeds from 
large vacuum pumps.
The cubes are modular
in the sense that they can be connected
to form larger vacuum 
chambers.

The system was 
designed
and
built to 
meet the requirements of 
the 
EDM$^3$ 
collaboration's 
measurement 
\cite{vutha2018orientation}
of 
the electron electric dipole moment
(eEDM).
This measurement uses polar 
molecules embedded in a
cryogenic
inert-gas
solid
\cite{corriveau2024matrix}.
The planned measurement requires
both an 
ultrapure 
solid
(which in turn requires an
extremely low vacuum pressure)
and exquisite control of magnetic
fields
(which in turn requires the 
vacuum system to be constructed
of nonmagnetic materials).

Although 
austenitic
grades of 
stainless steel 
are essentially nonmagnetic
when annealed, 
they  
exhibit 
ferro\-magnetic
properties when 
welded, 
cold-worked, 
machined, 
or exposed to extreme heat
\cite{tavares2002ferromagnetic}.
Aluminum
is clearly a better choice 
if no magnetization of 
the vacuum chamber can 
be tolerated.
Aluminum has the additional
advantages of being
lightweight and 
easily machined.

In this work
we 
consider a vacuum system
constructed from
two
of these
19-inch
aluminum
cubes
(detailed 
drawings are given in the 
Supplementary Materials).
We
calculate that 
a gas load
(from all sources:
leaks, 
permeation,
and 
outgassing)
of 
$Q_{\rm calc}$$<$$10^{-7}$~torr$\,$L/s
should be possible.
Large 
cryo\-pumps
with a total
pumping speed of 
$>$$10\,000$~L/s
would allow for 
a vacuum pressure of 
$P_{\rm calc}$$<$$10^{-11}$~torr.
As discussed in 
Section~\ref{sec:AlAdsorption},
the aluminum chamber walls
adsorb water vapor
(water vapor
is the dominant
gas load
for this system,
as discussed in 
Section~\ref{sec:OringOutgassing})
and act as a 
40\,000~L/s 
vacuum pump 
for several days
after being baked
(until 
the surfaces coverage 
starts leading to significant
outgassing).
This additional pumping 
allows for pressures 
of approaching 
$10^{-12}$~torr.

For our 
eEDM measurement,
the
cryogenic 
inert-gas
solid will 
be inside
a 
3-kelvin
copper box
(similar to
the one in 
Ref.~\cite{corriveau2024matrix}),
which
in turn will be
within
a 
40-kelvin
aluminum box.
This outer box 
has a surface area
of 
8\,000~cm$^2$,
and therefore 
will act
as an additional
cryo\-pump 
for water 
with a pumping speed
of 
$>$$100\,000$~L/s.
As this 
cryo\-pumping
speed is 
independent of coverage,
a pressure of 
$<$$10^{-12}$~torr 
can be sustained 
indefinitely.

Because the
3-kelvin 
copper 
box is 
very efficient 
at 
cryo\-pumping 
all gasses,
the vacuum pressure
inside of 
this box
will be much lower.
This box has  
a small
opening to allow 
a molecular
beam to enter.
A similar cryogenic 
system with a
small hole 
connecting to a
room-temperature
ultrahigh-vacuum
(UHV)
system
had a demonstrated 
pressure reduction
\cite{
fitzakerley2016electron}
of almost 
nine
orders of magnitude
(from
a 
UHV
pressure of
$>$$4$$\times$$10^{-8}$~torr
to
$<$$6$$\times$$10^{-17}$~torr,
as measured by the fact that
no 
antiproton 
annihilation was observed).
For the 
EDM$^3$
system, 
this factor may be even larger,
since in this system there is 
no
straight-line
path 
from the 
room-temperature 
vacuum system to the 
inside of the 
3-kelvin 
box,
making it necessary for
molecules to undergo one 
or more
bounces off of 
3-kelvin 
surfaces before entering the 
box.

A disadvantage 
of using aluminum
for constructing a
vacuum chamber 
is that standard
metal seals cannot be used
and therefore 
aluminum vacuum
systems are 
most often
sealed
with 
Viton 
O-rings. 
In 
Section~\ref{sec:leaks},
we discuss the challenges 
presented by leaks 
across 
Viton
O-ring
seals 
caused 
by scratches in 
the aluminum surfaces.
In 
Section~\ref{sec:permeation},
we discuss the further 
challenges presented by 
permeation of atmospheric
gases through 
Viton
O-rings.
In 
Section~\ref{sec:dualOring}
we discuss 
how both the challenges of
scratches and permeation
are overcome using
a
dual-O-ring 
design.
In 
Section~\ref{sec:surfaceOutgassing}
we discuss the advantages of 
aluminum over stainless steel
for surface 
outgassing
and in 
Section~\ref{sec:OringOutgassing}
we 
discuss the effect of 
outgassing 
from the 
Viton
O-rings,
which is the 
limiting factor for 
obtaining the lowest 
possible pressures.
In 
Section~\ref{sec:AlAdsorption}
we discuss
how adsorption of 
water
vapor
by the
baked 
walls of the 
aluminum 
chamber
(and,
for our
eEDM
measurement,
by cryogenic surfaces)
further 
improves the 
vacuum pressure.

Our initial
pump-down 
of this system
(for which
neither of these 
cryogenic boxes
was present)
achieved a vacuum 
pressure of 
$2$$\times$$10^{-9}$~torr
with an
1000-L/s
turbo\-pump.
A 
leak-up
rate of 
$2$$\times$$10^{-10}$~torr/s
when the system is not 
being actively pumped,
indicates 
an effective
outgassing 
rate of 
$5$$\times$$10^{-8}$~torr\,L/s,
as discussed 
in 
Section~\ref{sec:initialPump}.
This rate excludes water
vapor,
which is adsorbed onto the 
walls of the chamber,
but is the important rate
for our planned measurement,
as the very high 
sticking coefficient for
water will not allow it
to pass into
our 
3-kelvin 
cryogenic system.
The current system will be 
sufficient
to reach our target 
vacuum pressure 
(as discussed in 
Section~\ref{sec:concl}),
but it is also clear that
further baking 
(or 
prebaking)
of our
Viton
O-rings
would further lower
the 
outgassing 
rate.

\section{
leaks past 
Viton 
O-rings
\label{sec:leaks}
}
One concern for using 
Viton-sealed
aluminum vacuum 
chambers is 
that aluminum is not as 
resistant to scratches as 
stainless steel,
and 
small scratches in the aluminum
can allow air to leak
into the vacuum space.
To understand the effect of 
the scratches, 
we make the approximation 
that the scratches 
create leaks that 
are
equivalent to
cylindrical openings of radius
$R$
and 
length 
$\ell$,
where
$\ell$
is 
approximately the diameter 
of the 
O-ring
---
which is 
3~mm
for our 
Viton 
seals.
Such a cylindrical leak  
leads to a gas load 
of 
approximately
\begin{eqnarray}
\label{eq:leakRate}
&&
\!\!\!\!\!\!\!Q_{\rm scratch}=
\frac{100\,\rm L/s}
{{\rm cm}^{2}}
\frac{PR^3}{\ell}
\!+\!
\frac{8\,\rm L/s}
{{\rm cm}^{2}}
\frac{P R^4}{\ell \lambda}
\!=\!
\nonumber
\\
&&\frac{100\,\rm L/s}
{{\rm cm}^{2}}
\frac{PR^3}{\ell}
\!+\!
\frac{\rm 1500\, L/s}
{{\rm cm}^{3}\,{\rm torr}}
\frac{P^2 R^4}{\ell},
\end{eqnarray}
where 
$P$
is the pressure 
on one side of the 
O-ring
(with an assumed pressure
of zero on the other side)
and
$\lambda$ is the 
mean free path,
which has a value of
70~nm 
at atmospheric pressure
for 
room-temperature 
air
and varies linearly 
with 
$P$
at lower
pressures.
The first terms of 
Eq.~(\ref{eq:leakRate})
represents the gas load
from the conductance in 
the 
molecular-flow 
regime
($
\lambda/R>1
$),
the second term 
is the result in 
the 
no-slip
regime
($
\lambda/R<0.01
$),
and the sum 
gives an approximation
that spans both 
regimes.
(Although more accurate
formulas could be used in
the intermediate regime,
Eq.~(\ref{eq:leakRate})
gives the 
a
simple approximation
that is useful at the 
approximately
10\%
level,
which is, 
in any case,
higher accuracy than 
absolute pressures and 
gas loads can be
readily measured.)

From decades of experience with
aluminum-Viton 
seals, 
the authors find that a typical
leak into a vacuum system
from such a scratch 
(from
$P=1$~atmosphere)
has a 
gas load of 
the order of 
$10^{-5}$~torr$\,$L/s.
Indeed, 
in the initial 
pump-down
of our current
aluminum 
vacuum system
we had 
four such leaks 
(on the inner
O-rings
of our 
dual-O-ring
seals)
that 
ranged between
0.2
and
$5$$\times$$10^{-5}$~torr$\,$L/s.
Surprisingly to us, 
these four 
leaks sealed themselves
($Q_{\rm scratch}<$$2$$\times$$10^{-8}$~torr$\,$L/s)
after baking the system
at 
120$^\circ$C.
The other eight
seals 
used in this initial 
pump-down 
also
had leaks of 
$<$$2$$\times$$10^{-8}$~torr$\,$L/s.
Using the 
simple model of 
Eq.~(\ref{eq:leakRate})
for these leaks,
the implied radius 
$R$
ranges between
1.5 
and
3.5~microns
for the
four leaking 
seals
and 
less than
0.4~microns
for the 
nonleaking
seals. 

Aluminum-Viton
vacuum systems 
often have 
a dozen 
or more
O-ring
seals.
This set of seals
typically have a few
scratches on 
the
O-ring 
sealing surfaces
and therefore
often are limited 
by a gas load of 
the order of  
$3$$\times$$10^{-5}$~torr$\,$L/s.
At a pumping speed 
of 
10\,000~L/s,
this gas load leads to 
a limiting
vacuum pressure of 
3$\times$$10^{-9}$~torr.

\section{
permeation
through 
Viton
O-rings
\label{sec:permeation}
}

Permeation through 
Viton
O-rings 
is dominated by water
vapor
\cite{sefa2017comparison},
with over 
80\% 
of the gas load 
coming from water
vapor 
for 
air that has 
1.4\% 
absolute humidity
(a partial pressure
of 
10~torr).
The permeability for 
water 
vapor 
is approximately 
\cite{sefa2017comparison}
$K_{\rm water}=
3$$\times$$10^{-6}$~cm$^2$/s.
For our compressed 
Viton 
O-rings,
this permeation is through a 
distance of approximately 
$d=4$~mm
and over an area of approximately
$A
= 
2.5$~mm$\,$$\times$$140$~cm $=35$~cm$^2$
(a circumference 
of 
140~cm
and a compressed
cross section of 
approximately
$2.5$~mm$\,$$\times$$ \, 4$~mm),
leading to a gas flow of 
$
Q_{\rm water}
=
12\,
K_{\rm water} A P_{\rm water}/d
$
of approximately
$5$$\times$$10^{-5}$~torr$\,$L/s
(where the factor of 
12 
is due to the 
12
O-rings
in our system).
This gas load 
limits the 
vacuum pressure 
(for 
a pumping 
speed of 
10\,000~L/s)
to 
$5$$\times$$10^{-9}$~torr,
but unlike the scratches
discussed in the previous 
section, 
which can be avoided 
with sufficient care, 
this gas load is 
unavoidable for a 
single-O-ring
Viton-sealed
system.
As a result, 
most 
Viton-sealed
vacuum systems are 
limited to pressures 
of 
$5$$\times$$10^{-9}$~torr
or higher.
In the next section
we discuss a 
dual-O-ring
design that 
makes it possible to 
achieve much lower
pressures.

\section{
dual
O-ring
design
\label{sec:dualOring}
}

For our system,
we incorporate 
dual-O-ring seals
on all of our 
flanges.
Similar double seals have 
been used 
by others
(e.g,
Refs~\cite{de1967propertiesV,
park2017low,
cai2022nstx,
valente2021manufacturing}).
Our seals are
shown in 
cross section
in 
Fig.~\ref{fig:cubeDrawing}(c).
In this figure, the central 
$4.5$-mm-by-$4.5$~mm
channel acts as a secondary
vacuum system that is evacuated 
by a separate vacuum pump,
which is connected through the 
19-inch-diameter 
flange that covers the 
17-inch 
opening
(at the location indicated by
the dashed circle in 
Fig.~\ref{fig:cubeDrawing}(a),
and as described in 
detail in the 
Supplementary Materials).
The average conductance
of this channel
(averaged around its 
140-cm
circumference)
is approximately 
0.05~L/s.
With a gas load of 
approximately 
$10^{-5}
$~torr$\,$L/s
from permeation 
($
K_{\rm water} A P_{\rm water}/d
)$
and from scratches,
the average pressure 
in this channel 
will be 
approximately 
$P_{\rm channel}
=
2$$\times$$10^{-4}$~torr.

As a result of the low
pressure in the space
between the 
two 
O-rings,
the leakage 
through the inner
O-ring 
will be reduced by 
a factor of 
$P_{\rm channnel}
/
P_{\rm lab}$.
For 
permeation of
water 
vapor
($P_{\rm lab}^{\rm water}$$\approx
$10~torr),
this factor is 
$(2$$\times$$10^{-4}\,
{\rm torr})
/
(10\,{\rm torr})
\approx
2$$\times$$10^{-5}$,
which decreases the 
total gas
load from permeation 
from the 
$5$$\times$$10^{-5}$~torr$\,$L/s
given in 
Section~\ref{sec:permeation}
to
$1$$\times$$10^{-9}$~torr$\,$L/s.
Similarly, 
the 
$5$$\times$$10^{-9}$~torr 
limit of 
Section~\ref{sec:permeation}
reduces to 
$1$$\times$$10^{-13}$~torr.

The reduction factor for 
scratches is even larger:
$(2$$\times$$10^{-4}
\,{\rm torr})
/
(1\,{\rm atmosphere})
\approx
3$$\times$$10^{-7}$.
This factor 
reduces the 
$3$$\times$$10^{-5}$~torr$\,$L/s
gas load estimated for 
typical scratches to 
approximately
$1$$\times$$10^{-11}$~torr$\,$L/s.
Similarly, 
the
$3$$\times$$10^{-9}$~torr 
limit of 
Section~\ref{sec:leaks}
reduces to approximately 
$1$$\times$$10^{-15}$~torr.
The reduction factor is actually
even larger, 
since the gas load
given by 
Eq.~(\ref{eq:leakRate})
has a stronger than
proportional relationship
to pressure. 
Eq.~(\ref{eq:leakRate})
predicts that 
a scratch which would 
lead to a gas load of
$10^{-5}$~torr$\,$L/s
for a 
single-O-ring
system would lead to a 
gas load that is smaller by 
a factor of
$0.3\,
P_{\rm channnel}
/
P_{\rm lab}$
for our 
dual-O-ring
system,
leading to an even lower
gas load. 
As vacuum pressures of 
$10^{-15}$
are beyond the reach of this
aluminum-and-Viton
vacuum chamber, 
this additional factor is 
not important for 
typical-sized scratches.
However, 
for much larger scratches,
the factor becomes important.
Assuming scratches on both
the outer and inner 
O-rings 
that would
produce gas loads of 
$10^{-3}$~torr$\,$L/s
(100 
times larger than 
a typical scratch),
the intermediate pressure 
would increase to 
$P_{\rm channel}
=
2$$\times$$10^{-2}$~torr,
but the improvement factor due
to the 
dual-O-ring 
seal would be
$0.1\,
P_{\rm channnel}
/
P_{\rm lab}
\approx
3$$\times$$10^{-6}$.
As a result,
even if such large scratches 
were present at both seals, 
the gas load would be 
only
$3$$\times$$10^{-9}$~torr$\,$L/s
and 
the limiting 
pressure would still be  
$3$$\times$$10^{-13}$~torr.
Thus,
even scratches that are
much larger than those
typically encountered
lead to inconsequential 
increases in pressure.

The 
dual-O-ring
seal 
also has the 
advantage that leaks can 
easily be detected and 
measured by 
increasing the pressure 
(with air 
or 
argon)
in the intermediate
vacuum channel
between the two 
O-rings
(separtately for
each flange)
while monitoring the 
pressure in the main
vacuum system.
This method was used
here to find and measure
the four
leaks past the 
inner 
O-rings. 
To test the validity
of the approximation 
represented by 
Eq.~(\ref{eq:leakRate}),
we 
compared the gas load
into the main vacuum system
caused 
by introducing a gas with
a pressure of 
750~torr
and 
75~torr
into the 
channel between the 
two 
O-rings.
For the inner
O-ring that had 
a scratch causing a
$5$$\times$$10^{-5}$~torr$\,$L/s
leak,
the gas load reduced 
by a factor of
30
when the
channel pressure 
was reduced by a factor of 
10,
in agreement with 
the factor predicted by 
Eq.~(\ref{eq:leakRate}).

In summary,
the 
dual-O-ring
system
effectively
eliminates the problems
due to scratches and
permeation through
Viton,
and if these
two issues were the
only concerns,
vacuum pressures
of 
$10^{-13}$~torr 
would be possible.
In the next sections 
we discuss issues of 
outgassing, 
which dominate the 
actual vacuum pressure 
possible in 
this system.

\section{
outgassing 
from 
aluminum 
surfaces
\label{sec:OringOutgassing}
}

The initial 
outgassing 
rate for aluminum surfaces
exposed to air is
\cite{fedchak2021outgassing}
approximately
$10^{-8}$~torr$\,$L/s/cm$^2$,
with the 
outgassing 
being strongly dominated
by water 
vapor.
However, 
almost all of this
outgassing 
can be eliminated by baking 
to 
100$^\circ$C
for one day.
The residual 
outgassing 
is
dominated 
\cite{Moraw1974attainment}
by 
H$_2$
that is migrating out
of the bulk of the metal.
For 
stainless-steel, 
the 
H$_2$
outgassing 
rate 
is 
\cite{fedchak2021outgassing,
Moraw1974attainment}
approximately 
$7$$\times$$10^{-12}
$~torr$\,$L/s/cm$^2$.
If our chamber 
(with 
a surface area of 
approximately 
3~m$^2$)
were constructed from 
stainless steel,
the gas load would be
approximately
$2$$\times$$10^{-7}$~torr$\,$L/s.
With 
a pumping speed of
10\,000~L/s,
this would limit our 
vacuum pressure
to
$2$$\times$$10^{-11}$~torr.
Vacuum-firing
the chamber at a temperature
of 
950$^\circ$C
for one day
would reduce
\cite{fedchak2021outgassing}
the
H$_2$
gas load by two orders
of magnitude.

A similar reduction
is had by using 
aluminum in place of 
stainless steel
(without the need 
for firing).
Aluminum has
\cite{fedchak2021outgassing,
Moraw1974attainment}
an 
H$_2$
outgassing rate 
as low as
$4$$\times$$10^{-14}
$~torr$\,$L/s/cm$^2$,
which leads to a gas load of
$10^{-9}$~torr$\,$L/s.
A simulation of this surface outgassing 
for our
system with
two 
5\,000-L/s
cryopumps
using 
the 
CERN 
MolFlow+
software package,
leads to 
a
pressure of
$1$$\times$$10^{-13}$~torr
at the
center
position of 
each cube.

\section{
outgassing 
from 
Viton
O-rings
\label{sec:surfaceOutgassing}
}

In addition to gasses permeating
through 
Viton
O-rings,
one must
also be concerned with
continued outgassing from 
the Viton.
Viton 
O-rings 
exposed to air for a long time
have initial 
outgassing rates
\cite{de1966propertiesII}
of 
approximately
$10^{-6}$~torr$\,$L/s/cm$^2$.
However, 
properly baked out 
Viton
O-rings
after a day under vacuum
can have 
outgassing rates 
\cite{de1966propertiesI,
de1966propertiesIV,
hait1967application,
yoshimura1985differential}
of
$<$$10^{-10}$~torr$\,$L/s/cm$^2$.
In both cases, 
the 
outgassing is dominated
by water vapor.
For our system,
in which we have
twelve
140-cm-circumference
O-rings
with approximately
4~mm
of the 
O-ring 
being exposed to
the 
ultrahigh-vacuum 
chamber
(40\%
of the 
O-ring 
circumference 
outgasses
into the vacuum 
\cite{de1966propertiesII}),
the expected 
outgassing rate 
should be 
$<$$7$$\times$$10^{-8}$~torr$\,$L/s.
With 
10\,000~L/s
of pumping
(one
5\,000~L/s
cryro\-pump
at the bottom of
each cube),
this
gas load
leads to a pressure 
(again
using 
MolFlow+)
of
$<$$5$$\times$$10^{-12}$~torr
at the center of the cubes.

\section{
Adsorption of 
water 
vapor
onto the 
aluminum 
walls
\label{sec:AlAdsorption}
}

The sticking
coefficient 
for water vapor
on oxidized aluminum is 
\cite{krueger1972room}
approximately 
0.1.
This sticking 
coefficient 
reduces significantly as more water molecules adsorb
onto the surface
\cite{krueger1972room,
elam1998adsorption},
falling below 
the 
0.1
level for 
greater than 
approximately
$3$$\times$$10^{13}$
molecules per 
cm$^2$.
These tightly
bound water molecules
can have 
\cite{dobrozemsky2007residence}
multi-day
dwell
times at room temperature,
making the aluminum walls of our
freshly baked 
chamber an efficient 
getter for water molecules.
At a gas load of 
$<$$7$$\times$$10^{-8}$~torr$\,$L/s
(as predicted for our system in 
Section~\ref{sec:OringOutgassing}),
it would take
several days
to build up this 
density of molecules 
on the walls
of this chamber
after it has been thoroughly 
baked.
For 
this period of time, 
the walls
act as a getter
for water
vapor.
The total pumping speed
for this getter 
is 
40\,000~L/s
---
equivalent to the 
conductance through an
aperture of 
0.3-m$^2$
(the sticking 
coefficient of 
0.1
times the wall 
area of 
$\sim$3~m$^2$).
Using a
MolFlow+
simulation, 
we predict that 
vacuum pressures 
of
$<$$10^{-12}$~torr
would be possible
for the days during 
which the baked walls are
efficiently working 
as a getter for the 
water 
vapor 
that continues to outgas 
from the 
Viton
O-rings.

For our intended system
for the 
EDM$^3$ measurement,
which includes 
$8\,000$~cm$^2$ 
of surface kept
at 
40~kelvin,
the cold surface will
have a sticking
coefficient for 
water molecules of
unity
and therefore
act as a 
cryo\-pump
for water vapor
with a pumping 
speed of
$>$100\,000~L/s.
At this temperature,
the water molecules
will not 
desorb from the surface,
and the pumping speed
will be independent
of surface coverage,
allowing the pressure to 
be permanently lowered to
$<$$10^{-12}$~torr.

\section{
initial 
pump-down 
of our 
vacuum chamber
\label{sec:initialPump}
}

To date we have performed
only one 
pump-down 
using our 
aluminum-Viton
vacuum chamber.
A very simple 
procedure was used 
to clean the 
freshly-machined
aluminum parts:
a hand washing with 
warm soapy water,
a 
warm-water
rinse,
followed by 
several
wipes of all surfaces
using acetone 
and then
isopropanol
on 
Kimwipes 
or
swabs.
This
pump-down 
was performed by using two
connected cubic 
aluminum
chambers
(Fig.~\ref{fig:cubeDrawing}),
which were in turn connected to 
a 
stainless steel 
14-inch
CF
supercross.
The 
supercross
was
equipped with 
an
Edwards 
STP1003
turbo\-pump
(connected via an 
8-inch-inner-diameter
Kurt J Lesker
SG series 
gate valve),
a
Stanford Research Systems
RGA100
residual-gas
analyzer,
and an ion gauge.
The 
stainless-steel
system is
sealed with copper gaskets,
and is connected to the 
aluminum system using a 
titanium adapter.
We raised the 
temperature of the
entire aluminum system to 
over
120$^\circ$C
for 
87~hours.
The 
stainless-steel system was
baked less aggressively
(and at a less uniform
temperature)
to ensure that the 
turbo\-pump 
did not exceed its maximum allowed
baking temperature. 

The vacuum pressure 
obtained from this 
first 
pump-down 
was
$2$$\times$$10^{-9}$~torr
using a 
1000~L/s
turbo\-pump.
As expected from the discussion
in the previous sections,
the remaining gas load 
after baking
was dominated by water 
vapor.
Our 
residual-gas
analyzer 
(RGA)
indicates that 
approximately 
8\%
of the residual 
gas is
CO 
(or N$_2$)
(i.e., 
a mass of 
28 amu),
5\%
is 
CO$_2$,
and
1\%
is 
O$_2$.
No other 
gasses 
were evident from the 
RGA,
except for 
H$_2$,
which is discussed further below.
These relative 
partial pressures 
are more similar 
\cite{barton1965mass}
to those
those predicted from 
unbaked 
Viton
than for 
fully baked
Viton,
leading us to believe 
that more baking was
necessary in our case,
where the 
Viton
O-rings
were 
not 
prebaked,
but rather baked 
in-situ
(while compressed
in their grooves
and under vacuum).

The pressure of the system,
along with the known pumping 
speed of the 
turbo\-pump
of 
1000~L/s
(reduced by the pumping restrictions
above the pump, 
as calculated here 
using 
MolFlow+ 
to be 
700~L/s)
can be used to calculate a 
water-vapor 
gas load 
of 
$1.4\times10^{-6}$~torr\,L/s.
This gas load is an underestimate
given that the chamber 
walls also act as a getter for 
water vapor.
However, 
in our case, 
the water coverage on our 
surfaces is much larger
than the 
$3\times10^{13}$/cm$^2$ 
limit discussed in 
Section~\ref{sec:AlAdsorption},
and the sudden increase in 
the
water-vapor 
partial pressure 
when there is no active 
pumping 
(when the gate valve
in front of our
turbo\-pump
is closed)
indicates that the remaining
getter has a pumping speed
that is reduced to the level
of only 
approximately
100~L/s.
As a result, 
the revised estimate 
for our 
water-vapor gas load is
$1.6\times10^{-6}$~torr\,L/s. 
This result indicates
that the 
O-rings 
baking was not sufficient to 
get to the ultimate limit
of 
$<$$7$$\times$$10^{-8}$~Torr\,L/s 
that was discussed in 
Section~\ref{sec:OringOutgassing},
consistent with the conclusion 
deduced from the 
relative
RGA 
partial pressures.

The 
next-most-prominent
molecular mass of 
28 amu
likely 
includes a combination of 
N$_2$
(which is the 
second-most-important
gas for 
permeation through  
Viton)
and 
CO
(which can 
outgas from 
Viton 
due to chemical 
breakdown).
The gas load from these 
molecules
can 
be determined directly from 
a 
leak-up 
test,
as they
are not readily 
adsorbed
(or have an extremely short
dwell time after they
are adsorbed).
After closing our gate valve, 
the linear 
pressure increase 
was 
found to be 
$2$$\times$$10^{-10}$~torr/s.
This 
leak-up 
rate was 
found to be 
dominated by the increase
in the 
28-amu
molecules,
as expected
since the water molecules
continue
to be pumped 
(albeit with a smaller 
pumping speed)
by the
vacuum-chamber
walls.
This 
leak-up 
rate,
along with the 
MolFlow+
simulation of our 
vacuum system
indicates 
gas load
(from our 
12
O-rings)
of
$Q_{\rm eff}
$$=
$$5$$\times$$10^{-8}$~torr\,L/s.
Here, 
we define
$Q_{\rm eff}$
to be the effective
gas load for our system,
since the gas load from 
water vapor
(which is not included 
in 
$Q_{\rm eff}$)
is nearly irrelevant
for our vacuum system
given the anticipated 
large pumping speed
($>$$100\,000$-L/s)
that will be provided by
the cryogenic part of 
our apparatus
and 
given that the unity 
sticking coefficient for
water
at 
3~kelvin
will make any remaining 
water molecules unable to 
take the 
non-straight-line 
path into
our 
3-kelvin system.

Our 
RGA
was not well calibrated for
H$_2$.
Comparison of the 
changes 
(when heating or cooling 
parts of the vacuum system
and when closing the gate
valve in front of the 
turbo\-pump)
of
the 
H$_2$
partial-pressure,
the 
other partial pressures
measured by the RGA,
and the total pressure
measured by the ion gauge
allow for an approximate
calibration.
This calibration 
indicates that 
the partial pressure 
of 
H$_2$
is less than
$1$$\times$$10^{-10}$~torr,
which 
is consistent with 
the expected 
pressure from
outgassing
from the 
6\,500~cm$^2$
of unfired 
stainless
surfaces that we 
have present in our system.
It should be noted
that this 
H$_2$
gas load 
is 
not due to our aluminum
vacuum system, 
but rather the large 
stainless system that we 
used for the purpose 
of testing our aluminum 
system.

\section{conclusions 
and future prospects
\label{sec:concl}
}

In this work we present
an 
aluminum-and-Viton
vacuum chamber that
is designed to have
a total gas load
(dominated by 
outgassing 
of water
from 
the 
Viton
O-rings)
of 
$<$$10^{-7}$~torr\,L/s,
and that with
10\,000~L/s 
of vacuum pumping
(along with adsorption
of water onto the baked
chamber walls) 
will allow for a 
vacuum pressure in the 
extremely-high-vacuum
(XHV)
regime 
($\leq$$10^{-12}$~torr).
In our case, 
an additional 
$>$100\,000~L/s
of 
cryo\-pumping
on our cold apparatus
should allow for sustained
XHV 
vacuum pressures,
even after the 
coverage of the chamber
walls reduces their
adsorption efficiency.

An initial 
pump-down 
of the chamber
showed a somewhat higher
gas load of
$1.6$$\times$$10^{-6}$~torr\,L/s 
for water,
and 
$5$$\times$$10^{-8}$~torr\,L/s
for 
CO
and 
N$_2$.
Prebaking 
of the 
O-rings,
along with a more
extensive 
in-situ 
baking of the 
aluminum system
is expected to reduce
these rates to by over 
an order of magnitude
---
down to the rates found
by others 
\cite{de1966propertiesI,
de1966propertiesIV,
hait1967application,
yoshimura1985differential}
for 
baked 
Viton.
Even at the current
outgassing rate,
a 
10\,000~L/s
cryo\-pump,
along with 
the
$>$$100\,000$-L/s
cryo\-pumping from
the cryogenic portion of 
our apparatus, 
will allow for a 
pressure of 
$<$$10^{-11}$~torr.
This vacuum pressure is 
sufficient for our 
measurement given the fact
that we expect the pressure 
inside of the 
3-kelvin
volume 
to be lower by a factor of 
$10^9$ 
or more 
based on our previous
measurements of 
ultra-low 
pressures 
\cite{fitzakerley2016electron}
in a similar system.

\section*{Acknowledgments}
This work is supported by 
the 
Canada Foundation for Innovation,
the
Ontario Research Fund,
the 
Natural Sciences and Engineering Research Council
of Canada,
the 
Alfred P. Sloan Foundation,
the 
Gordon and Betty Moore Foundation, 
the 
Templeton 
Foundation
in conjunction with the 
Northwestern Center for Fundamental Physics, 
and 
York University. 

\bibliography{vacuum}

\begin{thebibliography}{20}%
\makeatletter
\providecommand \@ifxundefined [1]{%
 \@ifx{#1\undefined}
}%
\providecommand \@ifnum [1]{%
 \ifnum #1\expandafter \@firstoftwo
 \else \expandafter \@secondoftwo
 \fi
}%
\providecommand \@ifx [1]{%
 \ifx #1\expandafter \@firstoftwo
 \else \expandafter \@secondoftwo
 \fi
}%
\providecommand \natexlab [1]{#1}%
\providecommand \enquote  [1]{``#1''}%
\providecommand \bibnamefont  [1]{#1}%
\providecommand \bibfnamefont [1]{#1}%
\providecommand \citenamefont [1]{#1}%
\providecommand \href@noop [0]{\@secondoftwo}%
\providecommand \href [0]{\begingroup \@sanitize@url \@href}%
\providecommand \@href[1]{\@@startlink{#1}\@@href}%
\providecommand \@@href[1]{\endgroup#1\@@endlink}%
\providecommand \@sanitize@url [0]{\catcode `\\12\catcode `\$12\catcode `\&12\catcode `\#12\catcode `\^12\catcode `\_12\catcode `\%12\relax}%
\providecommand \@@startlink[1]{}%
\providecommand \@@endlink[0]{}%
\providecommand \url  [0]{\begingroup\@sanitize@url \@url }%
\providecommand \@url [1]{\endgroup\@href {#1}{\urlprefix }}%
\providecommand \urlprefix  [0]{URL }%
\providecommand \Eprint [0]{\href }%
\providecommand \doibase [0]{http://dx.doi.org/}%
\providecommand \selectlanguage [0]{\@gobble}%
\providecommand \bibinfo  [0]{\@secondoftwo}%
\providecommand \bibfield  [0]{\@secondoftwo}%
\providecommand \translation [1]{[#1]}%
\providecommand \BibitemOpen [0]{}%
\providecommand \bibitemStop [0]{}%
\providecommand \bibitemNoStop [0]{.\EOS\space}%
\providecommand \EOS [0]{\spacefactor3000\relax}%
\providecommand \BibitemShut  [1]{\csname bibitem#1\endcsname}%
\let\auto@bib@innerbib\@empty
\bibitem [{\citenamefont {Vutha}\ \emph {et~al.}(2018)\citenamefont {Vutha}, \citenamefont {Horbatsch},\ and\ \citenamefont {Hessels}}]{vutha2018orientation}%
  \BibitemOpen
  \bibfield  {author} {\bibinfo {author} {\bibfnamefont {AC}~\bibnamefont {Vutha}}, \bibinfo {author} {\bibfnamefont {M}~\bibnamefont {Horbatsch}}, \ and\ \bibinfo {author} {\bibfnamefont {EA}~\bibnamefont {Hessels}},\ }\bibfield  {title} {\enquote {\bibinfo {title} {Orientation-dependent hyperfine structure of polar molecules in a rare-gas matrix: A scheme for measuring the electron electric dipole moment},}\ }\href@noop {} {\bibfield  {journal} {\bibinfo  {journal} {Physical Review A}\ }\textbf {\bibinfo {volume} {98}},\ \bibinfo {pages} {032513} (\bibinfo {year} {2018})}\BibitemShut {NoStop}%
\bibitem [{\citenamefont {Corriveau}\ \emph {et~al.}(2024)\citenamefont {Corriveau}, \citenamefont {Lambo}, \citenamefont {Heinrich}, \citenamefont {Garcia}, \citenamefont {McCall}, \citenamefont {Yau}, \citenamefont {Chauhan}, \citenamefont {Koyanagi}, \citenamefont {Marsman}, \citenamefont {George}, \citenamefont {Storry}, \citenamefont {Horbatsch},\ and\ \citenamefont {Hessels}}]{corriveau2024matrix}%
  \BibitemOpen
  \bibfield  {author} {\bibinfo {author} {\bibfnamefont {Z}~\bibnamefont {Corriveau}}, \bibinfo {author} {\bibfnamefont {RL}~\bibnamefont {Lambo}}, \bibinfo {author} {\bibfnamefont {D}~\bibnamefont {Heinrich}}, \bibinfo {author} {\bibfnamefont {J~Perez}\ \bibnamefont {Garcia}}, \bibinfo {author} {\bibfnamefont {NT}~\bibnamefont {McCall}}, \bibinfo {author} {\bibfnamefont {H-M}\ \bibnamefont {Yau}}, \bibinfo {author} {\bibfnamefont {T}~\bibnamefont {Chauhan}}, \bibinfo {author} {\bibfnamefont {GK}~\bibnamefont {Koyanagi}}, \bibinfo {author} {\bibfnamefont {A}~\bibnamefont {Marsman}}, \bibinfo {author} {\bibfnamefont {MC}~\bibnamefont {George}}, \bibinfo {author} {\bibfnamefont {CH}~\bibnamefont {Storry}}, \bibinfo {author} {\bibfnamefont {M}~\bibnamefont {Horbatsch}}, \ and\ \bibinfo {author} {\bibfnamefont {EA}~\bibnamefont {Hessels}},\ }\bibfield  {title} {\enquote {\bibinfo {title} {Matrix isolated barium monofluoride: Assembling a sample of baf molecules for a measurement of the electron electric dipole
  moment},}\ }\href@noop {} {\bibfield  {journal} {\bibinfo  {journal} {arXiv preprint arXiv:2410.04591}\ } (\bibinfo {year} {2024})}\BibitemShut {NoStop}%
\bibitem [{\citenamefont {Tavares}\ \emph {et~al.}(2002)\citenamefont {Tavares}, \citenamefont {Da~Silva}, \citenamefont {Neto}, \citenamefont {Miraglia},\ and\ \citenamefont {Fruchart}}]{tavares2002ferromagnetic}%
  \BibitemOpen
  \bibfield  {author} {\bibinfo {author} {\bibfnamefont {SSM}\ \bibnamefont {Tavares}}, \bibinfo {author} {\bibfnamefont {MR}~\bibnamefont {Da~Silva}}, \bibinfo {author} {\bibfnamefont {JM}~\bibnamefont {Neto}}, \bibinfo {author} {\bibfnamefont {S}~\bibnamefont {Miraglia}}, \ and\ \bibinfo {author} {\bibfnamefont {D}~\bibnamefont {Fruchart}},\ }\bibfield  {title} {\enquote {\bibinfo {title} {{Ferromagnetic properties of cold rolled AISI 304L steel}},}\ }\href@noop {} {\bibfield  {journal} {\bibinfo  {journal} {Journal of Magnetism and Magnetic Materials}\ }\textbf {\bibinfo {volume} {242}},\ \bibinfo {pages} {1391} (\bibinfo {year} {2002})}\BibitemShut {NoStop}%
\bibitem [{\citenamefont {Fitzakerley}\ \emph {et~al.}(2016)\citenamefont {Fitzakerley}, \citenamefont {George}, \citenamefont {Hessels}, \citenamefont {Skinner}, \citenamefont {Storry}, \citenamefont {Weel}, \citenamefont {Gabrielse}, \citenamefont {Hamley}, \citenamefont {Jones}, \citenamefont {Marable}, \citenamefont {Tardiff}, \citenamefont {Grzonka}, \citenamefont {Oelert},\ and\ \citenamefont {Zielinski}}]{fitzakerley2016electron}%
  \BibitemOpen
  \bibfield  {author} {\bibinfo {author} {\bibfnamefont {DW}~\bibnamefont {Fitzakerley}}, \bibinfo {author} {\bibfnamefont {MC}~\bibnamefont {George}}, \bibinfo {author} {\bibfnamefont {EA}~\bibnamefont {Hessels}}, \bibinfo {author} {\bibfnamefont {TDG}\ \bibnamefont {Skinner}}, \bibinfo {author} {\bibfnamefont {CH}~\bibnamefont {Storry}}, \bibinfo {author} {\bibfnamefont {M}~\bibnamefont {Weel}}, \bibinfo {author} {\bibfnamefont {G}~\bibnamefont {Gabrielse}}, \bibinfo {author} {\bibfnamefont {CD}~\bibnamefont {Hamley}}, \bibinfo {author} {\bibfnamefont {N}~\bibnamefont {Jones}}, \bibinfo {author} {\bibfnamefont {K}~\bibnamefont {Marable}}, \bibinfo {author} {\bibfnamefont {E}~\bibnamefont {Tardiff}}, \bibinfo {author} {\bibfnamefont {D}~\bibnamefont {Grzonka}}, \bibinfo {author} {\bibfnamefont {W}~\bibnamefont {Oelert}}, \ and\ \bibinfo {author} {\bibfnamefont {M}~\bibnamefont {Zielinski}},\ }\bibfield  {title} {\enquote {\bibinfo {title} {Electron-cooled accumulation of 4$\times$$10^9$ positrons for
  production and storage of antihydrogen atoms},}\ }\href@noop {} {\bibfield  {journal} {\bibinfo  {journal} {Journal of Physics B: Atomic, Molecular and Optical Physics}\ }\textbf {\bibinfo {volume} {49}},\ \bibinfo {pages} {064001} (\bibinfo {year} {2016})}\BibitemShut {NoStop}%
\bibitem [{\citenamefont {Sefa}\ and\ \citenamefont {Setina}(2017)}]{sefa2017comparison}%
  \BibitemOpen
  \bibfield  {author} {\bibinfo {author} {\bibfnamefont {M}~\bibnamefont {Sefa}}\ and\ \bibinfo {author} {\bibfnamefont {J}~\bibnamefont {Setina}},\ }\bibfield  {title} {\enquote {\bibinfo {title} {{Comparison of permeation of atmospheric gases through Viton O-ring gaskets for different initial conditions}},}\ }\href@noop {} {\bibfield  {journal} {\bibinfo  {journal} {Journal of Vacuum Science \& Technology A}\ }\textbf {\bibinfo {volume} {35}},\ \bibinfo {pages} {041603} (\bibinfo {year} {2017})}\BibitemShut {NoStop}%
\bibitem [{\citenamefont {de~Csernatony}\ and\ \citenamefont {Crawley}(1967)}]{de1967propertiesV}%
  \BibitemOpen
  \bibfield  {author} {\bibinfo {author} {\bibfnamefont {L}~\bibnamefont {de~Csernatony}}\ and\ \bibinfo {author} {\bibfnamefont {DJ}~\bibnamefont {Crawley}},\ }\bibfield  {title} {\enquote {\bibinfo {title} {{The properties of Viton “A” elastomers. Part V. The practical application of Viton “A” seals in high vacuum}},}\ }\href@noop {} {\bibfield  {journal} {\bibinfo  {journal} {Vacuum}\ }\textbf {\bibinfo {volume} {17}},\ \bibinfo {pages} {551} (\bibinfo {year} {1967})}\BibitemShut {NoStop}%
\bibitem [{\citenamefont {Park}\ and\ \citenamefont {Cho}(2017)}]{park2017low}%
  \BibitemOpen
  \bibfield  {author} {\bibinfo {author} {\bibfnamefont {I-Y}\ \bibnamefont {Park}}\ and\ \bibinfo {author} {\bibfnamefont {B}~\bibnamefont {Cho}},\ }\bibfield  {title} {\enquote {\bibinfo {title} {{Low-carbon steel ultra-high-vacuum Schottky emitter electron gun with double O-rings for axis adjustment}},}\ }\href@noop {} {\bibfield  {journal} {\bibinfo  {journal} {Journal of Vacuum Science \& Technology A}\ }\textbf {\bibinfo {volume} {35}},\ \bibinfo {pages} {020604} (\bibinfo {year} {2017})}\BibitemShut {NoStop}%
\bibitem [{\citenamefont {Cai}\ \emph {et~al.}(2022)\citenamefont {Cai}, \citenamefont {Smith}, \citenamefont {Blanchard}, \citenamefont {Cai}, \citenamefont {Hennessy}, \citenamefont {Zhang},\ and\ \citenamefont {Loesser}}]{cai2022nstx}%
  \BibitemOpen
  \bibfield  {author} {\bibinfo {author} {\bibfnamefont {F}~\bibnamefont {Cai}}, \bibinfo {author} {\bibfnamefont {M}~\bibnamefont {Smith}}, \bibinfo {author} {\bibfnamefont {W}~\bibnamefont {Blanchard}}, \bibinfo {author} {\bibfnamefont {D}~\bibnamefont {Cai}}, \bibinfo {author} {\bibfnamefont {J}~\bibnamefont {Hennessy}}, \bibinfo {author} {\bibfnamefont {H}~\bibnamefont {Zhang}}, \ and\ \bibinfo {author} {\bibfnamefont {D}~\bibnamefont {Loesser}},\ }\bibfield  {title} {\enquote {\bibinfo {title} {{NSTX-U machine core vacuum seals upgrade design}},}\ }\href@noop {} {\bibfield  {journal} {\bibinfo  {journal} {IEEE Transactions on Plasma Science}\ }\textbf {\bibinfo {volume} {50}},\ \bibinfo {pages} {4274} (\bibinfo {year} {2022})}\BibitemShut {NoStop}%
\bibitem [{\citenamefont {Valente}\ \emph {et~al.}(2021)\citenamefont {Valente}, \citenamefont {Manfrin}, \citenamefont {Zaccaria}, \citenamefont {Barzon}, \citenamefont {Degli~Agostini}, \citenamefont {Fasolo}, \citenamefont {Rossetto}, \citenamefont {Tollin}, \citenamefont {Bolcato}, \citenamefont {Parma}, \citenamefont {Ruaro}, \citenamefont {Zanotto}, \citenamefont {Casa}, \citenamefont {Giupponi}, \citenamefont {Mico~Montava}, \citenamefont {Moreno}, \citenamefont {Rotti},\ and\ \citenamefont {Urbani}}]{valente2021manufacturing}%
  \BibitemOpen
  \bibfield  {author} {\bibinfo {author} {\bibfnamefont {M}~\bibnamefont {Valente}}, \bibinfo {author} {\bibfnamefont {S}~\bibnamefont {Manfrin}}, \bibinfo {author} {\bibfnamefont {P}~\bibnamefont {Zaccaria}}, \bibinfo {author} {\bibfnamefont {A}~\bibnamefont {Barzon}}, \bibinfo {author} {\bibfnamefont {F}~\bibnamefont {Degli~Agostini}}, \bibinfo {author} {\bibfnamefont {D}~\bibnamefont {Fasolo}}, \bibinfo {author} {\bibfnamefont {F}~\bibnamefont {Rossetto}}, \bibinfo {author} {\bibfnamefont {M}~\bibnamefont {Tollin}}, \bibinfo {author} {\bibfnamefont {D}~\bibnamefont {Bolcato}}, \bibinfo {author} {\bibfnamefont {A}~\bibnamefont {Parma}}, \bibinfo {author} {\bibfnamefont {D}~\bibnamefont {Ruaro}}, \bibinfo {author} {\bibfnamefont {M}~\bibnamefont {Zanotto}}, \bibinfo {author} {\bibfnamefont {M}~\bibnamefont {Casa}}, \bibinfo {author} {\bibfnamefont {M}~\bibnamefont {Giupponi}}, \bibinfo {author} {\bibfnamefont {G}~\bibnamefont {Mico~Montava}}, \bibinfo {author} {\bibfnamefont {JF}~\bibnamefont {Moreno}},
  \bibinfo {author} {\bibfnamefont {C}~\bibnamefont {Rotti}}, \ and\ \bibinfo {author} {\bibfnamefont {M}~\bibnamefont {Urbani}},\ }\bibfield  {title} {\enquote {\bibinfo {title} {{Manufacturing, on-site installation and acceptance test activities of the MITICA vacuum vessel}},}\ }\href@noop {} {\bibfield  {journal} {\bibinfo  {journal} {Fusion Engineering and Design}\ }\textbf {\bibinfo {volume} {169}},\ \bibinfo {pages} {112473} (\bibinfo {year} {2021})}\BibitemShut {NoStop}%
\bibitem [{\citenamefont {Fedchak}\ \emph {et~al.}(2021)\citenamefont {Fedchak}, \citenamefont {Scherschligt}, \citenamefont {Avdiaj}, \citenamefont {Barker}, \citenamefont {Eckel}, \citenamefont {Bowers}, \citenamefont {O’Connell},\ and\ \citenamefont {Henderson}}]{fedchak2021outgassing}%
  \BibitemOpen
  \bibfield  {author} {\bibinfo {author} {\bibfnamefont {JA}~\bibnamefont {Fedchak}}, \bibinfo {author} {\bibfnamefont {JK}~\bibnamefont {Scherschligt}}, \bibinfo {author} {\bibfnamefont {S}~\bibnamefont {Avdiaj}}, \bibinfo {author} {\bibfnamefont {DS}~\bibnamefont {Barker}}, \bibinfo {author} {\bibfnamefont {SP}~\bibnamefont {Eckel}}, \bibinfo {author} {\bibfnamefont {B}~\bibnamefont {Bowers}}, \bibinfo {author} {\bibfnamefont {S}~\bibnamefont {O’Connell}}, \ and\ \bibinfo {author} {\bibfnamefont {P}~\bibnamefont {Henderson}},\ }\bibfield  {title} {\enquote {\bibinfo {title} {Outgassing rate comparison of seven geometrically similar vacuum chambers of different materials and heat treatments},}\ }\href@noop {} {\bibfield  {journal} {\bibinfo  {journal} {Journal of Vacuum Science \& Technology B}\ }\textbf {\bibinfo {volume} {39}},\ \bibinfo {pages} {024201} (\bibinfo {year} {2021})}\BibitemShut {NoStop}%
\bibitem [{\citenamefont {Moraw}\ and\ \citenamefont {Dobrozemsky}(1974)}]{Moraw1974attainment}%
  \BibitemOpen
  \bibfield  {author} {\bibinfo {author} {\bibfnamefont {G}~\bibnamefont {Moraw}}\ and\ \bibinfo {author} {\bibfnamefont {R}~\bibnamefont {Dobrozemsky}},\ }\bibfield  {title} {\enquote {\bibinfo {title} {Attainment of outgassing rates below $10^{-13} $~torr$\,$liters/sec/cm$^2$ for aluminium and stainless steel after bakeout at moderate temperatures},}\ }\href@noop {} {\bibfield  {journal} {\bibinfo  {journal} {Japanese Journal of Applied Physics}\ }\textbf {\bibinfo {volume} {13}},\ \bibinfo {pages} {261} (\bibinfo {year} {1974})}\BibitemShut {NoStop}%
\bibitem [{\citenamefont {de~Csernatony}(1966{\natexlab{a}})}]{de1966propertiesII}%
  \BibitemOpen
  \bibfield  {author} {\bibinfo {author} {\bibfnamefont {L}~\bibnamefont {de~Csernatony}},\ }\bibfield  {title} {\enquote {\bibinfo {title} {{The properties of Viton “A” elastomers II. The influence of permeation, diffusion and solubility of gases on the gas emission rate from an O-ring used as an atmospheric seal or high vacuum immersed}},}\ }\href@noop {} {\bibfield  {journal} {\bibinfo  {journal} {Vacuum}\ }\textbf {\bibinfo {volume} {16}},\ \bibinfo {pages} {129} (\bibinfo {year} {1966}{\natexlab{a}})}\BibitemShut {NoStop}%
\bibitem [{\citenamefont {de~Csernatony}(1966{\natexlab{b}})}]{de1966propertiesI}%
  \BibitemOpen
  \bibfield  {author} {\bibinfo {author} {\bibfnamefont {L}~\bibnamefont {de~Csernatony}},\ }\bibfield  {title} {\enquote {\bibinfo {title} {{The properties of Viton “A” elastomers: I. Determination of the sorption and air solubility characteristics of Viton “A” and their effect in vacuum applications of the material}},}\ }\href@noop {} {\bibfield  {journal} {\bibinfo  {journal} {Vacuum}\ }\textbf {\bibinfo {volume} {16}},\ \bibinfo {pages} {13} (\bibinfo {year} {1966}{\natexlab{b}})}\BibitemShut {NoStop}%
\bibitem [{\citenamefont {de~Csernatony}(1966{\natexlab{c}})}]{de1966propertiesIV}%
  \BibitemOpen
  \bibfield  {author} {\bibinfo {author} {\bibfnamefont {L}~\bibnamefont {de~Csernatony}},\ }\bibfield  {title} {\enquote {\bibinfo {title} {{The properties of Viton “A” elastomers IV. The influence of solid-gas interaction at Viton “A” and stainless steel surfaces on gas evolution rates in high vacuum}},}\ }\href@noop {} {\bibfield  {journal} {\bibinfo  {journal} {Vacuum}\ }\textbf {\bibinfo {volume} {16}},\ \bibinfo {pages} {427} (\bibinfo {year} {1966}{\natexlab{c}})}\BibitemShut {NoStop}%
\bibitem [{\citenamefont {Hait}(1967)}]{hait1967application}%
  \BibitemOpen
  \bibfield  {author} {\bibinfo {author} {\bibfnamefont {PW}~\bibnamefont {Hait}},\ }\bibfield  {title} {\enquote {\bibinfo {title} {The application of polyimide to ultrahigh vacuum seals},}\ }\href@noop {} {\bibfield  {journal} {\bibinfo  {journal} {Vacuum}\ }\textbf {\bibinfo {volume} {17}},\ \bibinfo {pages} {547} (\bibinfo {year} {1967})}\BibitemShut {NoStop}%
\bibitem [{\citenamefont {Yoshimura}(1985)}]{yoshimura1985differential}%
  \BibitemOpen
  \bibfield  {author} {\bibinfo {author} {\bibfnamefont {N}~\bibnamefont {Yoshimura}},\ }\bibfield  {title} {\enquote {\bibinfo {title} {A differential pressure-rise method for measuring the net outgassing rates of a solid material and for estimating its characteristic values as a gas source},}\ }\href@noop {} {\bibfield  {journal} {\bibinfo  {journal} {Journal of Vacuum Science \& Technology A: Vacuum, Surfaces, and Films}\ }\textbf {\bibinfo {volume} {3}},\ \bibinfo {pages} {2177} (\bibinfo {year} {1985})}\BibitemShut {NoStop}%
\bibitem [{\citenamefont {Krueger}\ and\ \citenamefont {Pollack}(1972)}]{krueger1972room}%
  \BibitemOpen
  \bibfield  {author} {\bibinfo {author} {\bibfnamefont {WH}~\bibnamefont {Krueger}}\ and\ \bibinfo {author} {\bibfnamefont {SR}~\bibnamefont {Pollack}},\ }\bibfield  {title} {\enquote {\bibinfo {title} {{Room temperature adsorption of water by aluminum thin films}},}\ }\href@noop {} {\bibfield  {journal} {\bibinfo  {journal} {Surface Science}\ }\textbf {\bibinfo {volume} {30}},\ \bibinfo {pages} {280} (\bibinfo {year} {1972})}\BibitemShut {NoStop}%
\bibitem [{\citenamefont {Elam}\ \emph {et~al.}(1998)\citenamefont {Elam}, \citenamefont {Nelson}, \citenamefont {Cameron}, \citenamefont {Tolbert},\ and\ \citenamefont {George}}]{elam1998adsorption}%
  \BibitemOpen
  \bibfield  {author} {\bibinfo {author} {\bibfnamefont {JW}~\bibnamefont {Elam}}, \bibinfo {author} {\bibfnamefont {CE}~\bibnamefont {Nelson}}, \bibinfo {author} {\bibfnamefont {MA}~\bibnamefont {Cameron}}, \bibinfo {author} {\bibfnamefont {MA}~\bibnamefont {Tolbert}}, \ and\ \bibinfo {author} {\bibfnamefont {SM}~\bibnamefont {George}},\ }\bibfield  {title} {\enquote {\bibinfo {title} {{Adsorption of H$_2$O on a single-crystal $\alpha$-Al$_2$O$_3$ (0001) surface}},}\ }\href@noop {} {\bibfield  {journal} {\bibinfo  {journal} {The Journal of Physical Chemistry B}\ }\textbf {\bibinfo {volume} {102}},\ \bibinfo {pages} {7008} (\bibinfo {year} {1998})}\BibitemShut {NoStop}%
\bibitem [{\citenamefont {Dobrozemsky}\ \emph {et~al.}(2007)\citenamefont {Dobrozemsky}, \citenamefont {Menhart},\ and\ \citenamefont {Buchtela}}]{dobrozemsky2007residence}%
  \BibitemOpen
  \bibfield  {author} {\bibinfo {author} {\bibfnamefont {R}~\bibnamefont {Dobrozemsky}}, \bibinfo {author} {\bibfnamefont {S}~\bibnamefont {Menhart}}, \ and\ \bibinfo {author} {\bibfnamefont {K}~\bibnamefont {Buchtela}},\ }\bibfield  {title} {\enquote {\bibinfo {title} {{Residence times of water molecules on stainless steel and aluminum surfaces in vacuum and atmosphere}},}\ }\href@noop {} {\bibfield  {journal} {\bibinfo  {journal} {Journal of Vacuum Science \& Technology A}\ }\textbf {\bibinfo {volume} {25}},\ \bibinfo {pages} {551} (\bibinfo {year} {2007})}\BibitemShut {NoStop}%
\bibitem [{\citenamefont {Barton}\ and\ \citenamefont {Govier}(1965)}]{barton1965mass}%
  \BibitemOpen
  \bibfield  {author} {\bibinfo {author} {\bibfnamefont {RS}~\bibnamefont {Barton}}\ and\ \bibinfo {author} {\bibfnamefont {RP}~\bibnamefont {Govier}},\ }\bibfield  {title} {\enquote {\bibinfo {title} {{A mass spectrometric study of the outgassing of some elastomers and plastics}},}\ }\href@noop {} {\bibfield  {journal} {\bibinfo  {journal} {Journal of Vacuum Science and Technology}\ }\textbf {\bibinfo {volume} {2}},\ \bibinfo {pages} {113} (\bibinfo {year} {1965})}\BibitemShut {NoStop}%
\end{thebibliography}%

\newpage

\section*{Supplementary Materials
\label{sec:supplMat}}

\begin{figure}[h]
\includegraphics[width=\linewidth]
{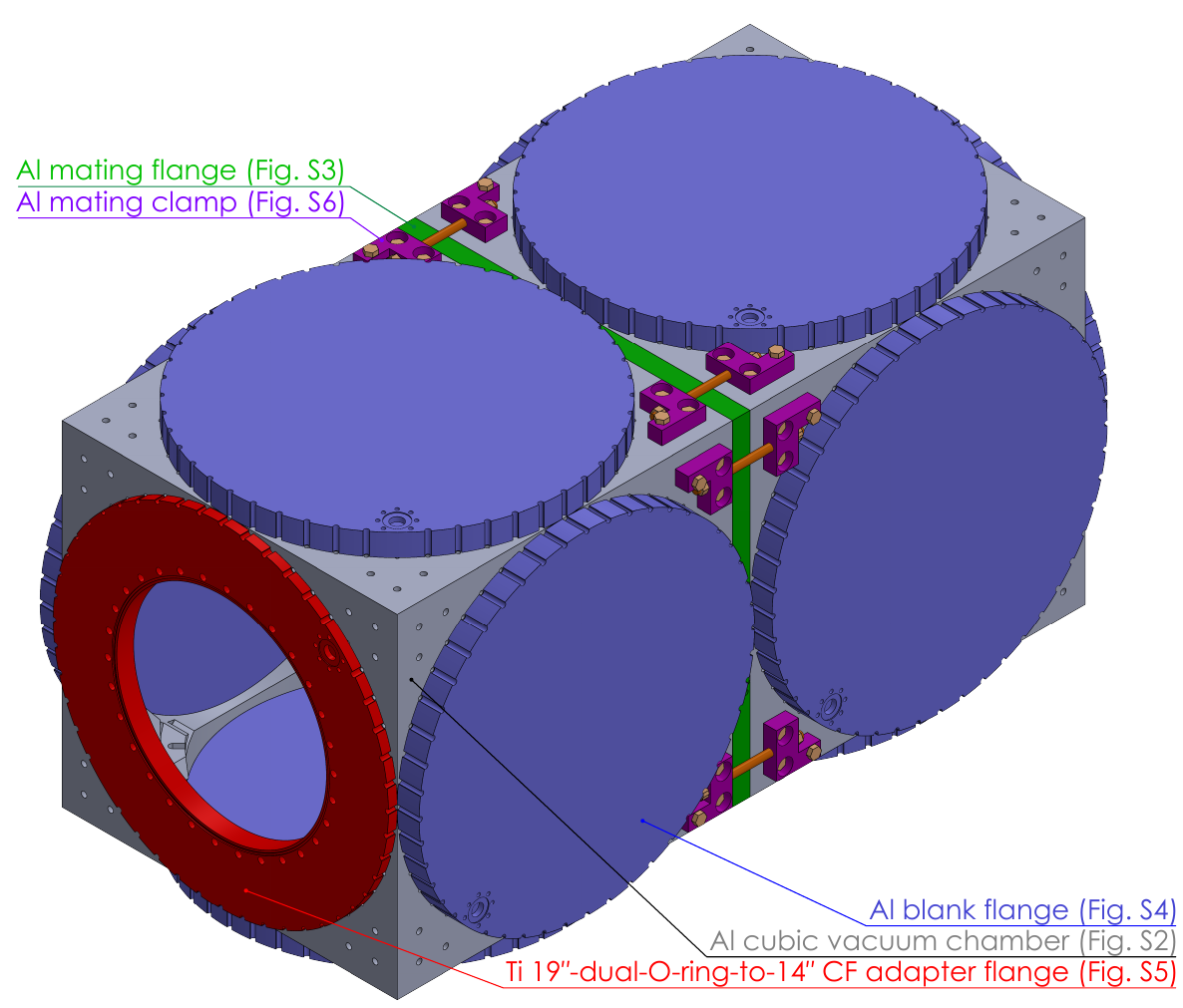}
\renewcommand{\thefigure}{S1}
\caption{
(color online)
An overview
of the 
vacuum system 
used for the 
current measurement.
The individual 
components
and their details
are shown in 
Figs.~\ref{fig:cubeDetails}
through
\ref{fig:matingClamp}.
}
\label{fig:chamberOverview}
\end{figure}

\begin{figure*}[h]
\includegraphics[width=0.9\linewidth]
{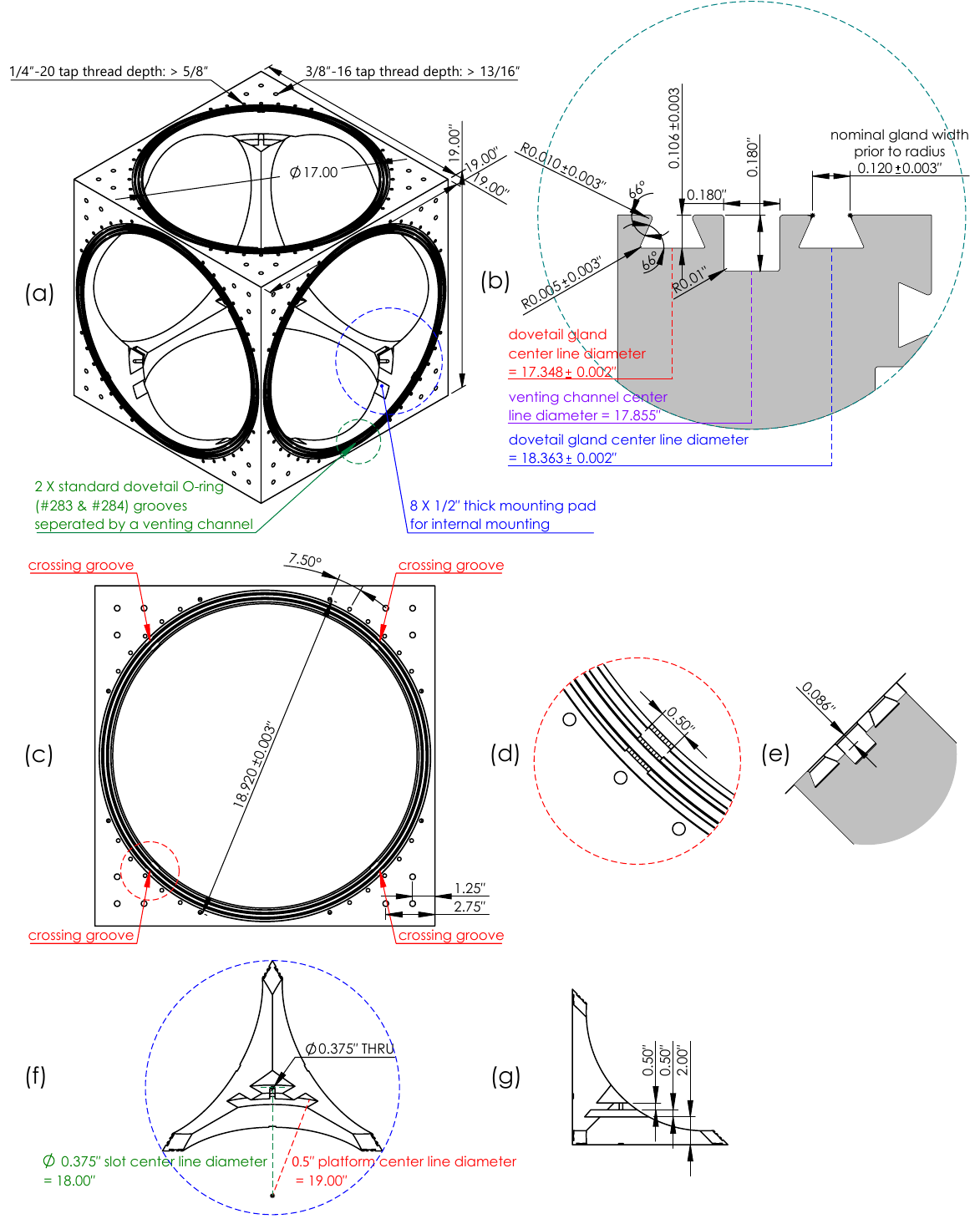}
\renewcommand{\thefigure}{S2}
\caption{
(color online)
Details for the 
aluminum cubic
vacuum chamber.
This chamber was machined
from a cube of 
5000-series
aluminum
(Clinton Aluminum
20 PLT 5000 mold).
The machining tolerances 
are 
0.005 inches,
except for where specified
on the drawings.
Surface roughness is 
specified at 
$\leq$125~microinches 
roughness average 
(Ra), 
except for the bottom
and sides 
of 
the
dovetail
O-ring 
grooves,
which are specified 
at 
$\leq$16
and
$\leq$32~microinches 
Ra,
respectively.
The 
dovetail 
O-ring
grooves must be machined on 
a lathe.
Panel~(a)
shows the overview of the 
cubic chamber.
Panel~(b) 
shows the 
cross-sectional view of the 
dovetail 
O-ring 
grooves and the 
venting channel
between the 
two 
O-ring 
grooves.
Panel~(c)
shows the 
positions of the 
threaded holes 
and of the
crossing grooves
that connnect the 
venting channel to the
O-ring 
grooves.
Panel~(d) 
and
(e)
show
the 
profile 
and 
depth,
respectively,
of 
the crossing grooves.
The 
crossing grooves 
must be machined 
before the dovetail grooves to
reduce burrs
in the 
O-ring
grooves.
Panels~(f)
and 
(g)
give a 
three-dimensional 
view
and 
diagonal 
cross-section
of the 
internal 
mounting pads.
}
\label{fig:cubeDetails}
\end{figure*}

\begin{figure*}[h]
\includegraphics[width=\linewidth]
{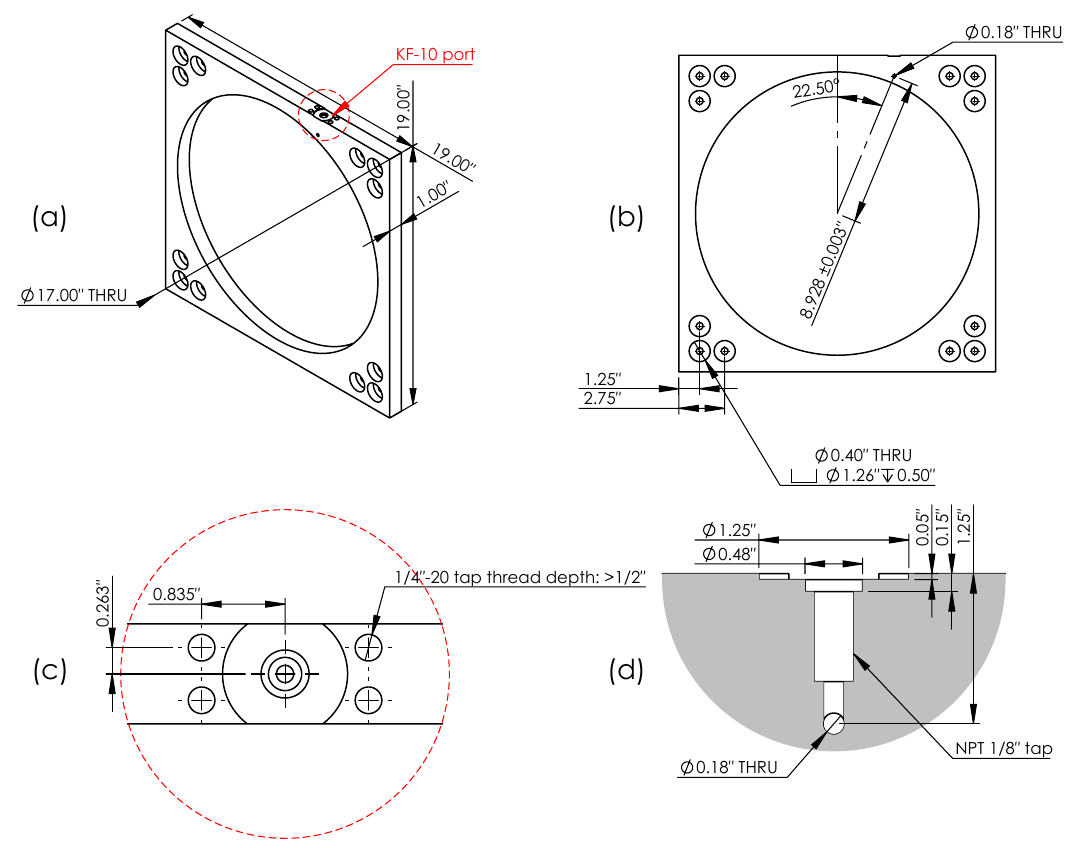}
\renewcommand{\thefigure}{S3}
\caption{
(color online)
Details for the 
6061
aluminum mating 
flange.
The machining tolerances 
are 
0.005 inches,
except for where specified
on the drawings.
Surface roughness is 
specified at 
$\leq$125~microinches 
Ra, 
except for the 
O-ring 
surfaces
(both the 
dual 
O-rings
and the 
KF-10
O-ring)
which are specified 
at 
$\leq$16~microinches 
Ra.
Panel~(a)
shows the 
three-dimensional
view of the flange.
Panel~(b)
gives the locations
for the holes 
and 
KF-10
port.
Panel~(c)
shows the top view 
of the 
KF-10
port
and
panel~(d)
shows the 
cross-sectional
side view.
}
\label{fig:matingFlange}
\end{figure*}

\begin{figure*}[h]
\includegraphics[width=\linewidth]
{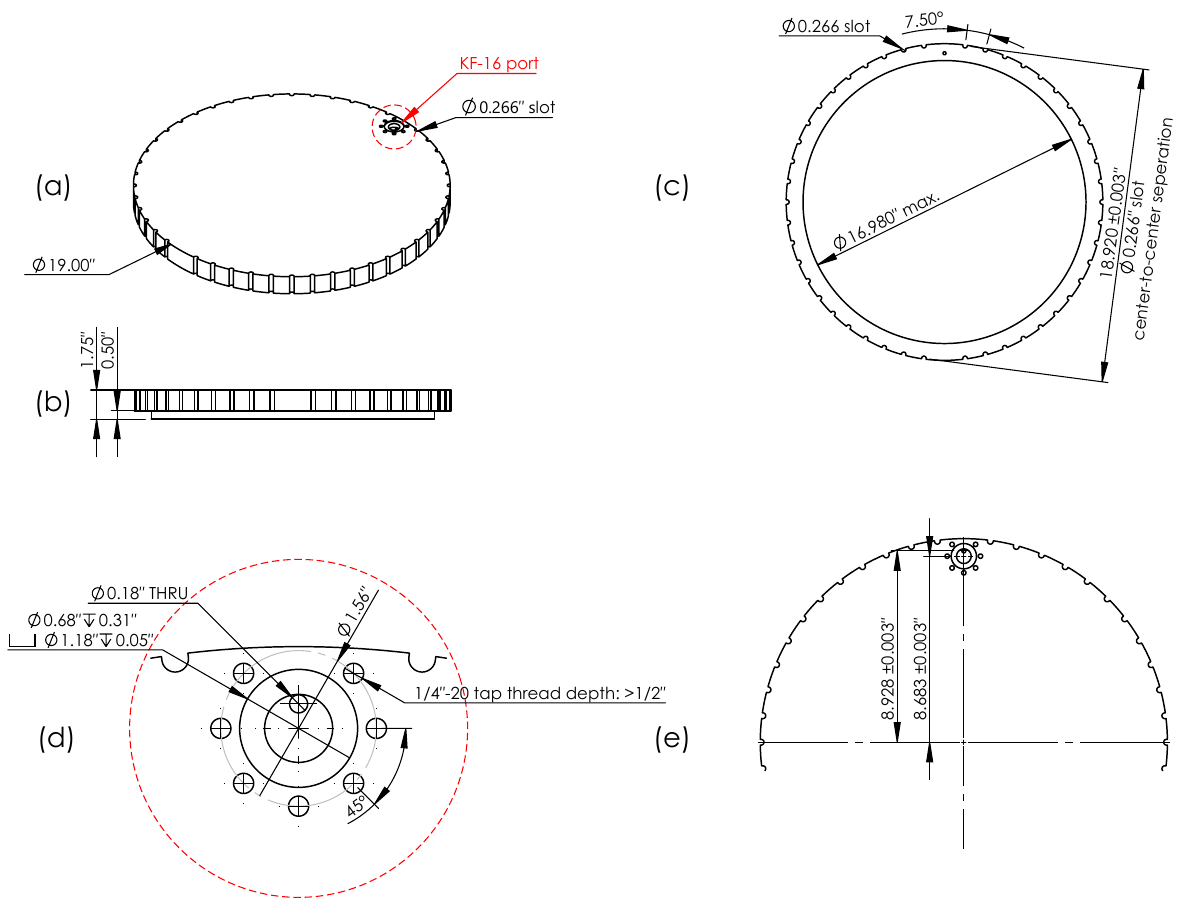}
\renewcommand{\thefigure}{S4}
\caption{
(color online)
Details for the 
6061
aluminum blank 
flange.
The machining tolerances 
are 
0.005 inches,
except for where specified
on the drawings.
Surface roughness is 
specified at 
$\leq$125~microinches 
Ra, 
except for the 
O-ring 
surfaces
(both the 
dual
O-ring
surface on the bottom of 
the flange
and the 
KF-16
O-ring
surface on the top)
which are specified 
at 
$\leq$16~microinches 
Ra.
O-ring
surface for dual 
O-rings
must be 
machined
on a lathe.
The 
flange should not be warped
(with top and 
bottom surfaces parallel to 
each other, 
with a maximum 
surface deviation from a 
plane of 0.003 inches).
Panel~(a)
gives a 
three-dimensional 
view of the flange.
Panel~(b)
gives a side view
and
panel~(c) 
gives a bottom view.
Panels~(d)
and 
(e)
give the details 
and position for 
the 
KF-16
port.
}
\label{fig:blankFlange}
\end{figure*}

\begin{figure*}[h]
\includegraphics[width=\linewidth]
{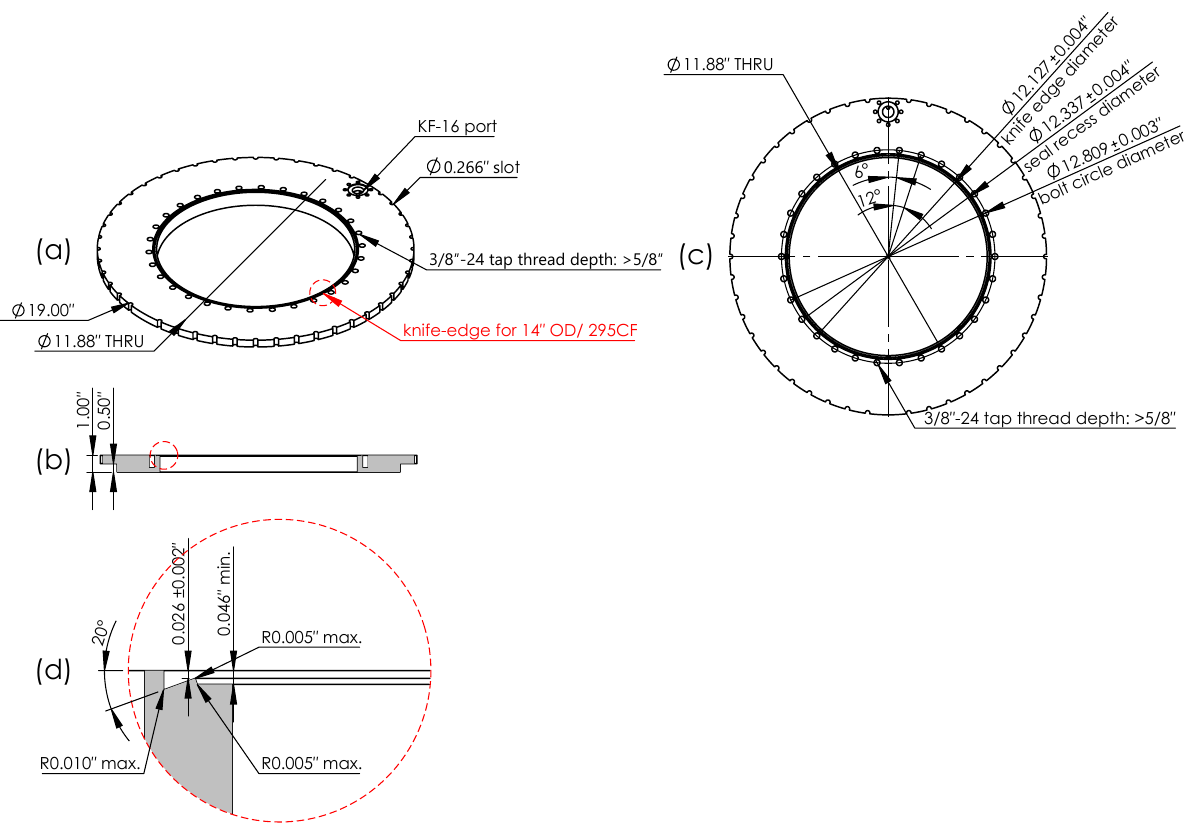}
\renewcommand{\thefigure}{S5}
\caption{
(color online)
Details for the 
Grade-5
titanium
19-inch-dual-O-ring--to--14-inch-CF
adapter flange.
Grade~5 
was required,
as the 
less-expensive 
Grade~2
is too soft to hold 
a durable knife edge.
The machining tolerances 
are 
0.005 inches,
except for where specified
on the drawings.
Surface roughness is 
specified at 
$\leq$125~microinches 
Ra, 
except for the 
O-ring 
surfaces
and 
knife edge,
which are specified 
at 
$\leq$16~microinches 
Ra.
Panel~(a)
gives the 
three-dimensional 
view.
Panels~(b)
and
(c) 
show the details for 
a standard 
14-inch
CF
knife-edge
seal.
}
\label{fig:adapterFlange}
\end{figure*}

\begin{figure*}[h]
\includegraphics[width=0.7\linewidth]
{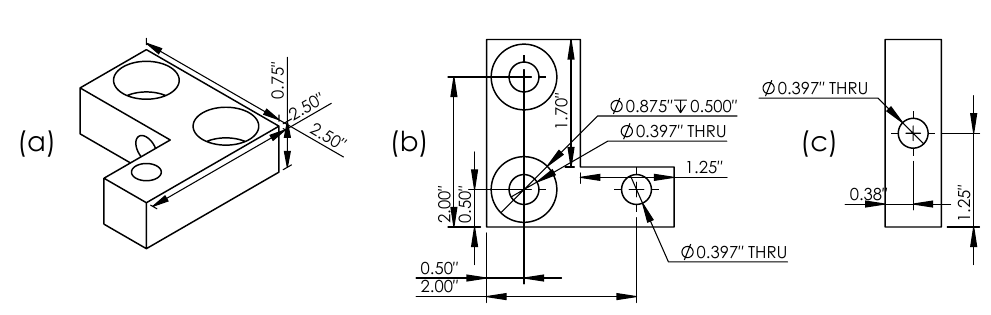}
\renewcommand{\thefigure}{S6}
\caption{
(color online)
Details for the 
6061
aluminum
17-inch-to-14-inch-CF
mating clamp.
The machining tolerances 
are 
0.005 inches
and the
surface roughness is 
specified at 
$\leq$125~microinches 
Ra.
}
\label{fig:matingClamp}
\end{figure*}

In this section we present 
detailed drawings of the 
components of our
system.
Fig.~\ref{fig:chamberOverview}
gives an overview
of a chamber
constructed from 
two 
aluminum cubes.
Fig.~\ref{fig:cubeDetails}
gives the detailed drawings 
for the aluminum cubic chamber.
Fig.~\ref{fig:matingFlange}
gives the detailed drawings 
for the flange used to mate
two cubes.
Fig.~\ref{fig:blankFlange}
gives the detailed drawings for
the blank flange used to 
cover the 
17-inch
cube openings.
Fig.~\ref{fig:adapterFlange}
gives the detailed drawings
for the titanium
adapter flange which
is used to adapt down
to a standard 
14-inch 
Conflat 
(CF)
metal-sealed
flange.
Fig.~\ref{fig:matingClamp}
gives the drawings for 
the clamp used to mate two
cubes.
Only brass bolts and nuts
are used as fasteners 
for constructing
the structure of 
Fig.~\ref{fig:chamberOverview},
and titanium
helicoil 
inserts are
used to reinforce the 
1/4"-20
aluminum 
threads
on the cubic
chamber.
No grease was used on 
the 
Viton
O-rings.

\end{document}